\RequirePackage{lineno}
\documentclass[aps,prc,twocolumn, floatfix]{revtex4}
\usepackage{graphicx}
\usepackage{dcolumn}
\usepackage{csquotes}
\usepackage{bm}
\usepackage{amsmath}

\begin{document}


\title{$\Omega^-$ production as a probe of equation of state of dense matter near the QCD phase transition in relativistic heavy-ion collisions}

\author{Zhi-Min Wu$^{1,2}$}
\author{Gao-Chan Yong$^{1,2}$}
\email[]{yonggaochan@impcas.ac.cn}

\affiliation{
$^1$School of Nuclear Science and Technology, University of Chinese Academy of Sciences, Beijing 100049, China\\
$^2$Institute of Modern Physics, Chinese Academy of Sciences, Lanzhou 730000, China
}

\begin{abstract}

The production of doubly strange hyperon $\Xi^-$ and trebly strange hyperon $\Omega^-$ in relativistic Au+Au collisions at $\sqrt{s_{NN}}$ = 4.2 GeV is explored based on a relativistic transport model that is interweaved with hadronic mean-field potentials for heavy-ion collisions. Upon comparison, it appears that relative to the double strangeness observable $\Xi^-$, the yield and collective flows of the triple strange $\Omega^-$ exhibit a higher sensitivity to the equation of state (EoS) of dense matter. This characteristic makes the $\Omega^-$ an essential observable for studying the properties of densely formed matter in relativistic heavy-ion collisions.

\end{abstract}

\maketitle

\section{Introduction}
Since the discovery of the strongly-coupled Quark-Gluon Plasma (sQGP) in ultra-high-energy heavy-ion collisions in 2005 \cite{star1,star2,star3,star4}, there has been a significant interest in this fascinating form of matter. People are particularly interested in understanding its properties and the phase transition process from hadronic gas to QGP. Although the phase transition from QGP to the hadronic phase, in the high temperature and vanishing net-baryon density region of the QCD phase diagram, is understood to be a smooth crossover thanks to high energy heavy-ion collisions and lattice QCD calculations \cite{ntr1}, the overall structure of the nuclear matter phase diagram is still less understood, particularly in the high baryon density region. The EoS of this new form of matter and the understanding of the structure of QCD phase holds significant relevance in nuclear physics and astrophysics \cite{pr1,sci1}. The EoS plays a key role in dictating the structure of compact stars \cite {akm1998,nature2020, liapj2020, hu2021, kojo2021, ran2021, xie2021} and also influences gravitational-wave (GW) emission \cite{gw2019, gw2018, gw20182}. Moreover, examining the characteristics of strongly interacting matter results in an exceptional understanding of the universe \cite{ann2006}, and the nucleosynthesis occurring during the presupernova evolution of large stars \cite{bethe1990,latt2001}. Given the crucial significance of investigating the boundary and structure of the QCD phase diagram, both theorists and experimenters are presently directing their attention to such projects. In fact, global territorial laboratories like CBM and HADES at GSI/FAIR, DHS and DS at J-PARC-HI, BM@N and MPD at NICA, FXT and BES-II at RHIC-STAR, CEE at HIAF will cover varied energy regions and afford unique opportunities to explore the QCD phase structure and the EoS of nuclear matter at high baryon densities \cite{tet2019}.

Recent investigations suggest that the EoS of dense matter tends to stiffen with the increase in baryon density. However, when the density surpasses a certain threshold, a softer EoS is required \cite{dmy2022,jan2022,wu2023}. This startling phenomenon warrants further exploration. The lowest order perturbative QCD theory posits that the production of $s\bar{s}$ quark pairs primarily results from the annihilation of light quark-antiquark pairs and collisions of gluons \cite{rafe1982}. As a result, one can expect a rise in strangeness abundance in high-energy heavy-ion collisions. The double strangeness $\Xi^-$ has been recently scrutinized as effective sensors for the nuclear EoS at high densities \cite{yong2021plb,yong2022prc}. Such in-depth analyses have highlighted its significant sensitivity compared to single strangeness. Consequently, it is intriguing to determine whether the triple strangeness hyperon $\Omega^-$ exhibits an enhanced sensitivity to the EoS of dense nuclear matter. Presumably, it originates predominantly from a higher density region and experiences minimal interactions with other baryons post production. Therefore, in this study, we investigate the production of $\Omega^-$, including its yields and collective flows, in Au+Au collisions at $\sqrt{S_{NN}}$=4.2 GeV. The results demonstrate that both its yields and collective flows show outstanding sensitivity to the EoS of dense matter. This finding offers a promising way to probe the EoS of dense nuclear matter in relativistic heavy-ion collisions and aids in elucidating the phase transition boundary and the onset of deconfinement in dense hadronic matter.

\section{The AMPT-HC model}
The Multi-Phase Transport model, known as AMPT, is a Monte Carlo transport model predominantly used for simulating heavy-ion collisions at relativistic energies. Its ability to incorporate both partonic and hadronic interactions makes it highly relevant for studying the complex dynamics within these collisions. The model's structure includes four key components: varying initial conditions, partonic interactions, processes that convert the partonic phase to the hadronic phase (hadronization), and hadronic interactions \cite{AMPT2005}. Over time, the AMPT model has been widely applied to study heavy-ion collisions at both RHIC and LHC energies \cite{nst2021}. Suitable for collision energies ranging from several GeV to several TeV, the model has recently been broadened through an additional execution version designed for pure hadron cascade with hadronic mean-field potentials \cite{yong2021plb}. This is known as the AMPT-HC mode, which is akin to an updated version of a relativistic transport (ART) model \cite{art95}. The AMPT-HC mode offers a detailed account of heavy-ion collisions at low beam energies near or prior to the onset of the QCD phase transition.

Within the hadron cascade section of the AMPT model, the production and secondary interactions for particles such as $\pi$, $\rho$, $\omega$, $\eta$, $K$, $K^*$, $\phi$, $n$, $p$, $\Delta$, $N^*(1440)$, $N^*(1535)$, $\Lambda$, $\Sigma$, $\Xi$, $\Omega$ and deuteron are incorporated. Other particles having PYTHIA flavor codes such as $D$, $D_s$, $J/\Psi$, $B$, $\Upsilon$, among others, can be produced but do not participate in secondary interactions \cite{AMPT2005,deu2009}. In the pure hadron cascade mode AMPT-HC, the initialization of nucleons' coordinates and momentum within the projectile and target nuclei is facilitated through the Woods-Saxon nucleon density distribution and local Thomas-Fermi approximation. Nucleons, baryon resonances, $K$, $\Lambda$, $\Sigma$, $\Xi$, $\Omega$, and their corresponding antiparticles undergo hadron mean-field potentials utilizing the test-particle method and the quark counting rule \cite{qcr74,qcr01}. Channels involving transitions like $Y+Y \rightleftharpoons N+\Xi$ and $Y+N \rightleftharpoons N+\Xi+K$ ($Y=\Lambda$ or $\Sigma$) are considered to accommodate $\Omega$ production via processes like $K+\Xi \rightleftharpoons \pi+ \Omega$ at lower beam energies. Given limited reliable knowledge about the nucleon potential at higher momenta and densities (for example, above 1 GeV/c and 3 times saturation density studied here), a simple, Skyrme-type parametrization mean field $U(\rho)=\alpha\frac{\rho}{\rho_0}+\beta(\frac{\rho}{\rho_0})^\gamma$ \cite{yong2021plb,guo2021} is used with incompressibility coefficients of $K_0$ = 200, 400 MeV for soft and stiff EoSs, respectively. However, the stiff EoS with $K_0$ = 400 MeV at high baryon density might violate causality. The primary aim of our current study is to identify an observable sensitive to the EoS by varying the $K_0$, not to constrain the EoS itself. Therefore, the physical implications would still be valid, even if the stiff EoS used may violate causality.
While not intending to constrain the EoS, we adopt this approach for research convenience. Additionally, recent research suggests that a softer EoS may be more applicable to dense nuclear matter. Therefore, some observables were calculated using an extremely soft EoS with $K_0$ = 100 MeV for comparison.


\section{Results and Discussions}
The impact parameter in heavy-ion collisions significantly influences the collision process's dynamics. Near-central collisions see a higher number of participating nucleons, resulting in more nucleon-nucleon interactions and subsequently a region of collision with higher energy density, promoting the production of strangeness. Semi-central collisions, however, display more intricate dynamics, encompassing a range of collective motions. In this research, we are specifically focusing on Au+Au collisions happening at $\sqrt{S_{NN}}$ = 4.2 GeV. An impact parameter of b = 2 fm is employed for the near-central collision scenario, whereas an impact parameter of b = 7 fm is applied for the semi-central collision scenario.

\begin{figure}[t]
\centering
\vspace{0.0cm}
\includegraphics[ width=0.45\textwidth]{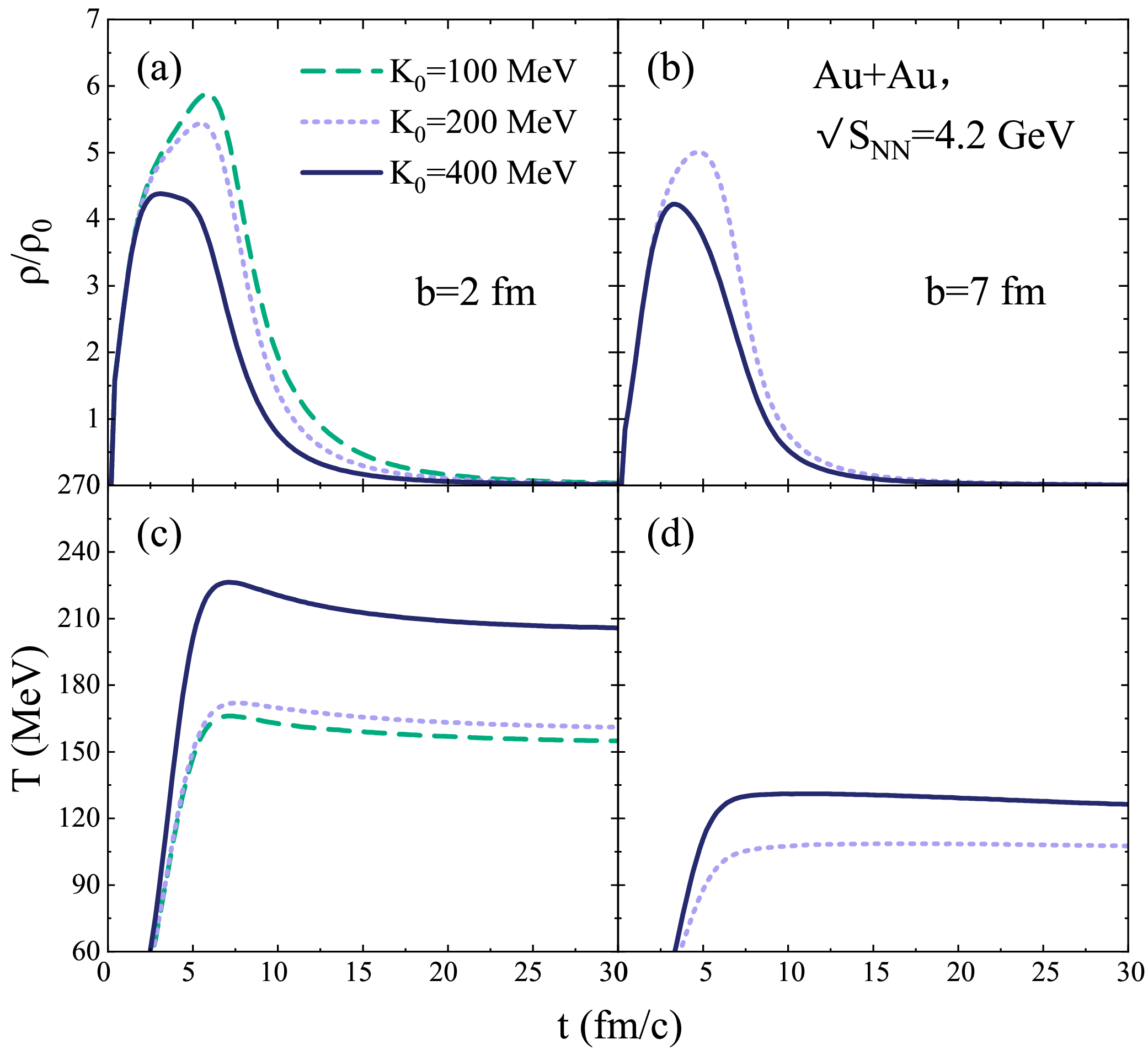}
\caption{Top: The evolution of compression baryon density at the central cell with soft ($K_0$ = 200, 100 MeV) and stiff ($K_0$ = 400 MeV) EoSs in both central and semi-central Au+Au collisions at $\sqrt{S_{NN}}$ = 4.2 GeV. Bottom: Evolution of the corresponding temperatures with differing EoSs.}\label{density}
\end{figure}
Investigating the properties of compressed dense matter requires initially determining the maximum compression baryon density reached in heavy-ion collisions. Shown in the top panels of Figure~\ref{density} is the time development of compression baryon density for both soft and stiff EoSs in central and semi-central Au+Au collisions at $\sqrt{S_{NN}}$=4.2 GeV. It's clear to see that the soft EoS yields a greater nuclear compression, particularly for central collisions. With central Au+Au collisions, the maximum baryon density attains roughly 4.3-5.8$\rho/\rho_{0}$, contingent on the EoS chosen. For semi-central Au+Au collisions, the baryon densities attained are generally lesser than those in central collisions. In addition, compressed matter endures longer with the soft EoS compared to the stiff one. Varied compressions in heavy-ion collisions with either soft or stiff EoSs can lead to different strangeness productions and diverse collective movements of the strangeness. Displayed in the bottom panels of Figure~\ref{density} are the developments of corresponding temperatures with varying EoSs. The temperature $T$ of the reaction system is approximated via $<2/3*E_{kin}>$, or $<2/3*1.5*E^{kin}_{trans}>$, where E$^{kin}_{trans}$ is representative of a particle's transverse kinetic energy. Before reaching maximum compression, we observe a quick rise in temperature which then plateaus. The stiff EoS leads to a considerably higher temperature than the soft EoS. According to the Gibbs Free Energy formula $G = H - TS$, the increase in temperature signifies a decrease in Gibbs Free Energy. Since the stiff EoS results in greater energy conversion to thermal energy, leaving less energy for new particle generation. One could therefore expect lesser strangeness production with the stiff EoS.

\begin{figure}[t]
\centering
\vspace{0.0cm}
\includegraphics[ width=0.45\textwidth]{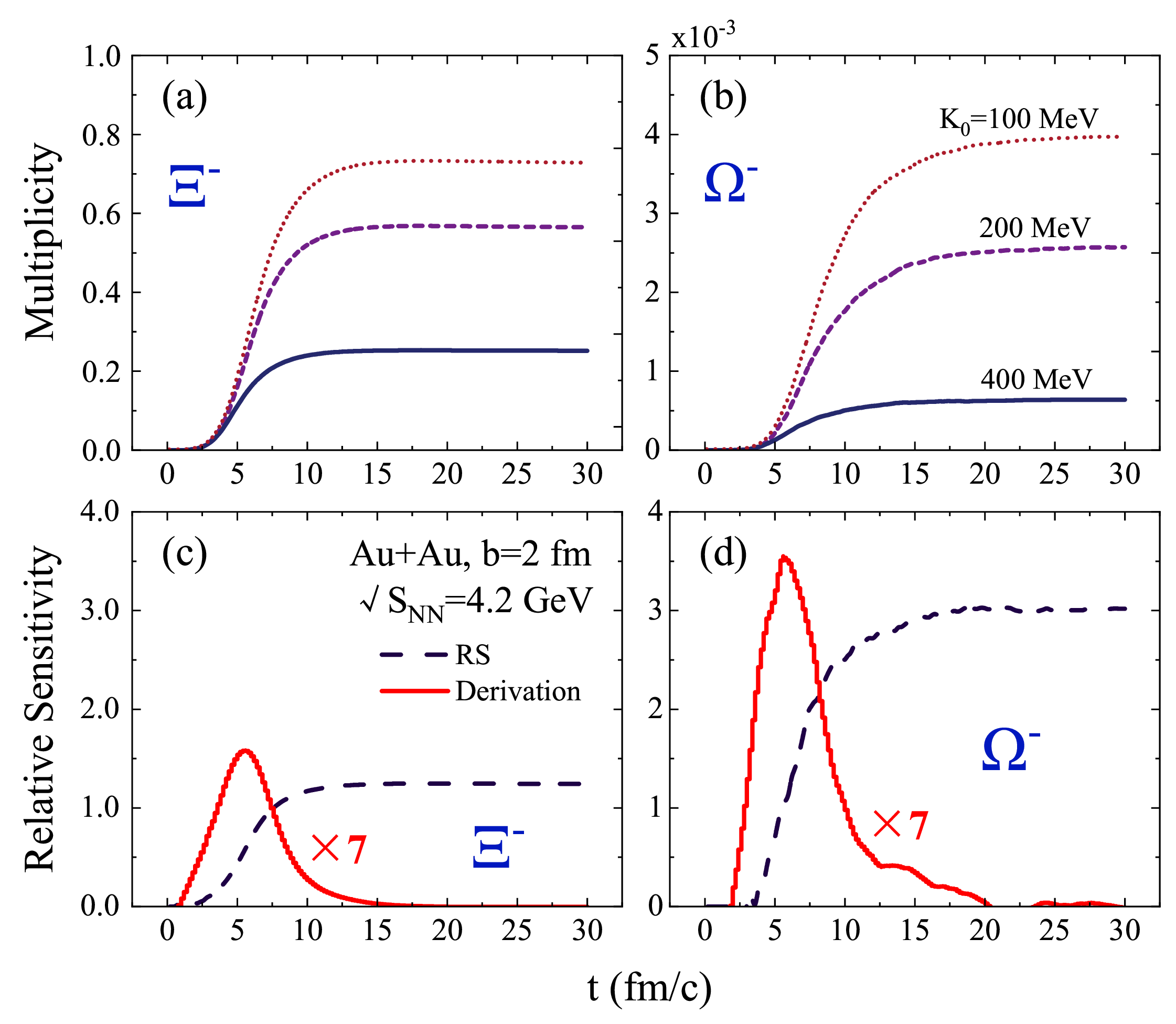}
\caption{Top: The multiplicities of $\Xi^-$ (left panel) and $\Omega^-$ (right panel) over time using soft and stiff EoSs, respectively, in central Au+Au collisions at $\sqrt{S_{NN}}$ = 4.2 GeV. Bottom: The corresponding relative sensitivity (black dot dash lines) and its derivative (red solid lines) for $\Xi^-$ and $\Omega^-$ over time.}
\label{sensi}
\end{figure}
Figure~\ref{sensi} depicts the production rates of $\Xi^-$ and $\Omega^-$ in central Au+Au collisions at $\sqrt{S_{NN}}$=4.2 GeV over time using soft and stiff EoSs respectively. The top panels of Figure~\ref{sensi} reveal that both $\Xi^-$ and $\Omega^-$ production rates display a sensitivity to the EoS, with the triple strangeness $\Omega^-$ being particularly sensitive. The $\Xi^-$ production rate is roughly 250 times higher than that of $\Omega^-$. The greater nuclear compression and Gibbs Free Energy linked with the soft EoS leads to a higher strangeness generation. When compared to Figure~\ref{density}, a subtle difference can be detected in the production rates of double strangeness $\Xi^-$ versus triple strangeness $\Omega^-$. The production of $\Xi^-$ commences at around 2.5 fm/c when the central compression density hits about 4.5 $\rho/\rho_0$, while $\Omega^-$ production begins near 5 fm/c, aligning with the central maximum compression density of approximately 5.5 $\rho/\rho_0$ (for soft EoS $K_0$ = 200 MeV). This suggests that $\Omega^-$ is created in denser matter than $\Xi^-$, thus, it should be more sensitive to the EoS. The EoS sensitivity of $\Omega^-$ here is about twice that of $\Xi^-$. As shown in Figures~\ref{density} and~\ref{sensi}, we have opted to plot observables for the extremely soft EoS ($K_0$=100 MeV) under specific scenarios. Our goal is not to limit the EoS, but rather to identify an observable that shows sensitivity to the EoS. It is clear from the aforementioned figures that the three distinct EoSs illustrate similar physics. However, presenting three different EoSs can lead to crowded figures, resulting in reduced clarity. Therefore, we've refrained from presenting results for the extremely soft EoS in later computations. A comparison between a soft and hard EoS sufficiently illustrates the EoS sensitivity of the observables.

In order to verify the results obtained in this study, we utilized the PHSD transport model (with partonic degrees-of-freedom turned off) \cite{mor2019} for an in-depth exploration. According to the calculations made with the PHSD model, the EoS sensitivity of $\Omega^-$ is about twice that of $\Xi^-$, which aligns with our findings. The bottom panels of Figure~\ref{sensi} illustrate the relative sensitivity of $\Xi^-$ and $\Omega^-$ over time. The relative sensitivity $RS$ is designated as $RS = (M(t)_{soft} - M(t)_{stiff})/M(t)_{stiff}$, where $M(t)_{soft}$ and $M(t)_{stiff}$ denote the multiplicities encountered with soft ($K_0$ = 200 MeV) and stiff ($K_0$ = 400 MeV) EoSs respectively, at a particular point in time. Figure~\ref{sensi} makes it obvious that the relative sensitivity of $\Omega^-$ is approximately 2.5 times that of $\Xi^-$. The derivative of $RS$ measures the sensitivity's rate of change and sheds light on how strangeness production is influenced by the EoS. Panels (c) and (d) show that both $\Xi^-$ and $\Omega^-$ particles display the highest sensitivity increase around 5 fm/c, corresponding to the highest compression point. The significant peak of the $RS$ derivative for $\Omega^-$ implies that $\Omega^-$ strangeness more acutely probes the maximum compression point than $\Xi^-$, thus exhibiting high sensitivity to the EoS. On the other hand, the prominent production channel of $\Omega^-$ is $\bar{K}+\Xi$ $\rightleftharpoons$ $\pi+\Omega$, which merges the sensitivities of single strangeness $\bar{K}$ and double strangeness $\Xi$. As a result, the triple strangeness $\Omega^-$ exhibits even higher sensitivity to the EoS than the double strangeness $\Xi^-$.

As illustrated in Figure~\ref{density}, the soft Equation of State (EoS) is associated with a higher degree of nuclear compression. This increase in nuclear matter compression results in a larger number of particle-particle inelastic collisions, thereby leading to an elevated production of strangenesses. Conversely, the soft EoS also results in a lower temperature of compressed matter. Since less energy is converted into thermal energy, an increased amount of Gibbs Free Energy is accessible for the generation of new particles. Bridging these two factors, it is plausible to anticipate a greater strangeness production associated with the soft EoS. In contrast to the non-strange pion meson acting as a probe of the EoS, once produced, the triple strangeness $\Omega^-$ is infrequently absorbed into the surrounding matter. The lack of final-state interactions enables it to be an effective probe for the EoS of dense matter generated in heavy-ion collisions. Also, the strangeness is produced at the stage of nuclear compression \cite{yong2021plb}, thereby it carries more extensive information about the properties of compressed matter than the pion.

\begin{figure}[t]
\centering
\vspace{0.0cm}
\includegraphics[ width=0.45\textwidth]{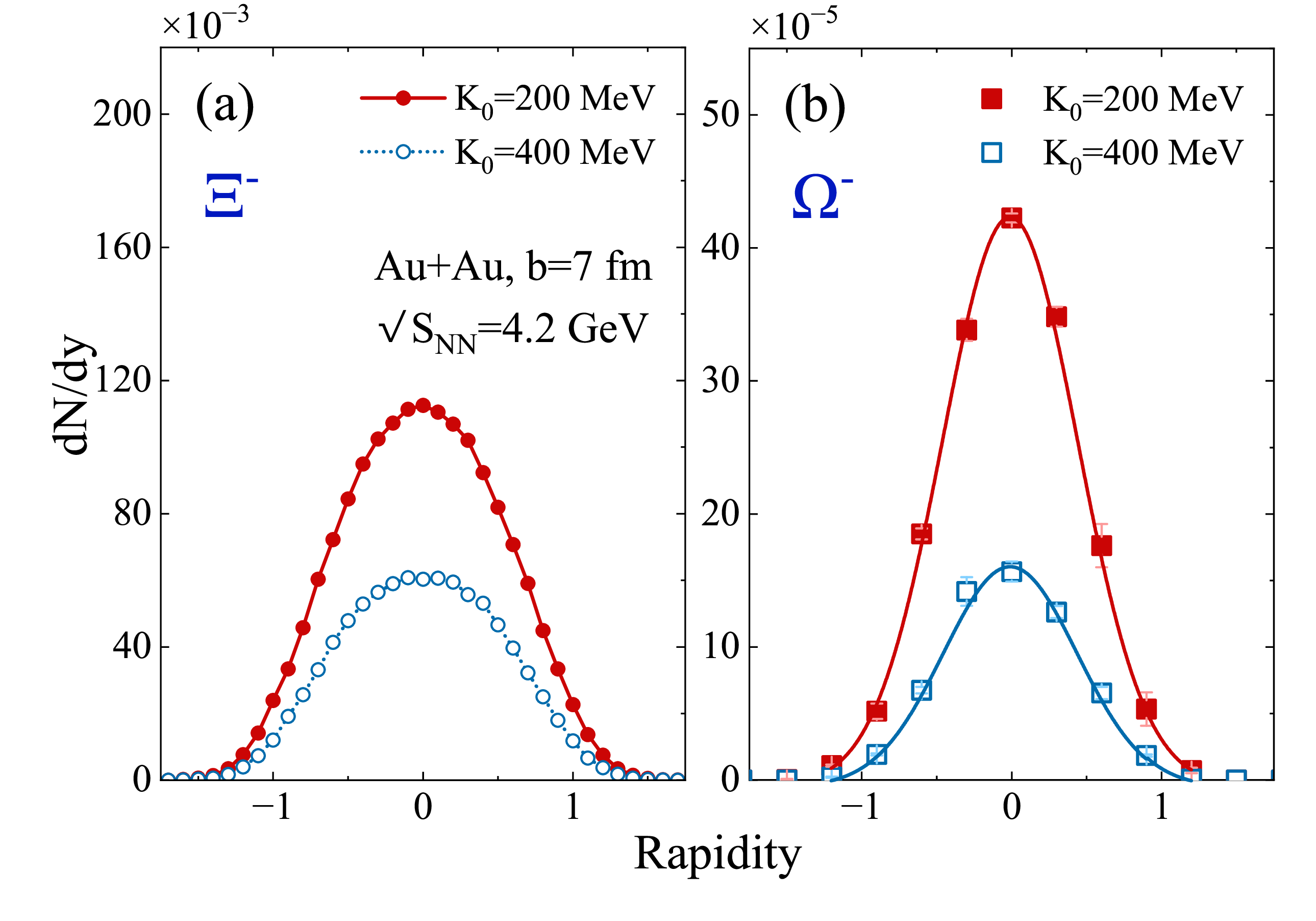}
\caption{Rapidity distributions of $\Xi^-$ (solid and hollow circles in left panel) and $\Omega^-$ (solid and hollow squares in right panel) with different EoSs in semi-central Au+Au collisions at $\sqrt{S_{NN}}$=4.2 GeV.} \label{ydis}
\end{figure}
The distribution of rapidity/transverse momentum is deemed vital for comprehending the dynamics of the collision system \cite{udd2012}. Prior research implies that the particle rapidity distribution is related to the spatial distribution of early collisions. Consequently, the central rapidity region, typically characterized as the spatial zone around $z \approx 0$, can be utilized to estimate the energy density of the initial state \cite{bjo1983}. Figure~\ref{ydis} depicts the rapidity distributions of $\Xi^-$ and $\Omega^-$ particles with varying EoSs in semi-central Au+Au collisions at $\sqrt{S_{NN}}$=4.2 GeV. It is observable that the EoS profoundly impacts the rapidity distributions of $\Xi^-$ and $\Omega^-$ particles, particularly at the mid-rapidity region. A soft EoS induces larger compression and an increased number of particle collisions, leading to a higher production of mid-rapidity strangenesses. It is also noted that the influences of EoS are more pronounced for mid-rapidity triple strangeness $\Omega^-$ than for double strangeness $\Xi^-$.

The correlation of collective motion in relativistic heavy-ion collisions to the bulk properties of the resulting matter is profound. It is essential to decode the information regarding the initial conditions, dynamics, degrees of freedom, and the equation of states. This can be achieved by representing the variation in the ultimate particles' distribution as the Fourier series:
\begin{flalign}
&\ E\frac{d^3N}{d^3p} = \frac{1}{2\pi} \frac{d^2N}{p_tdp_tdy}(1+\sum_{n=1}^{\infty}2v_ncos(n(\phi - \Psi))), &\label{eq3}
\end{flalign}
The notations $p_t$, $y$, $\phi$ and $\Psi$ symbolize the transverse momentum, rapidity, azimuthal angle, and the event plane angle of the particle respectively \cite{sorge1997prl,sorge1997plb,oll1992}. Directed flow and elliptic flow are embodied by the initial two Fourier expansion coefficients
$v_1 = \left \langle \frac{p_x}{p_t} \right \rangle $ and $v_2 = \left \langle \frac{p_x^2-p_y^2}{p_x^2+p_y^2} \right \rangle$ correspondingly. These flows are a subject of extensive study given their high sensitivities to the properties of dense matter present in heavy-ion collisions provide crucial insights \cite{hung1995,stein2014,nara2016,dani2002,dani1998,kru1985}.

\begin{figure}[t]
\centering
\vspace{0.0cm}
\includegraphics[width=0.45\textwidth]{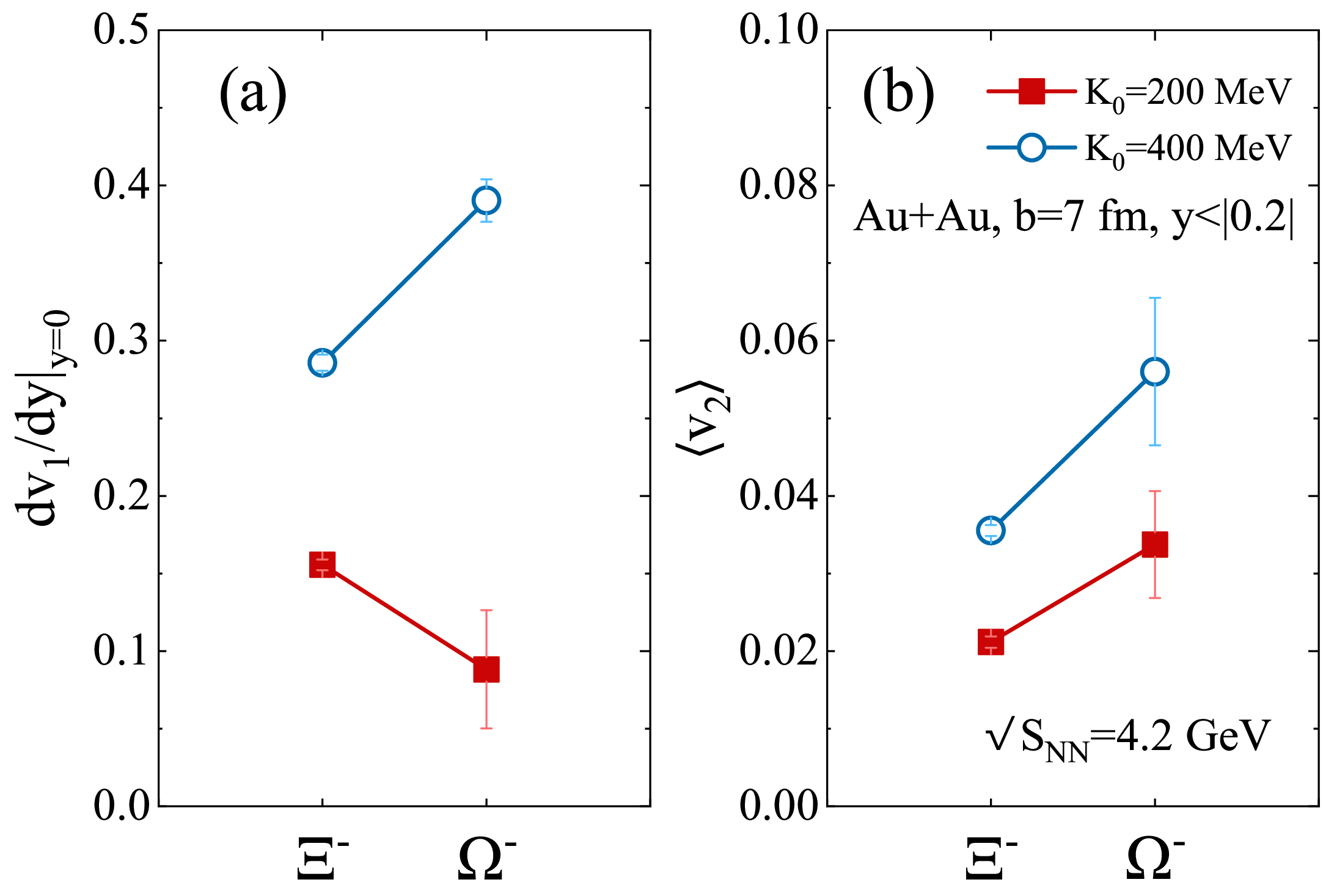}
\caption{$dv_1/dy|_{y=0}$ (left panel) and $\left \langle v_2 \right \rangle$ (right panel) of strangeness with soft and stiff EoSs in semi-central Au+Au collisions at $\sqrt{S_{NN}}$=4.2 GeV.} \label{v1v2y}
\end{figure}
\begin{figure}[t!]
\centering
\vspace{0.0cm}
\includegraphics[ width=0.45\textwidth]{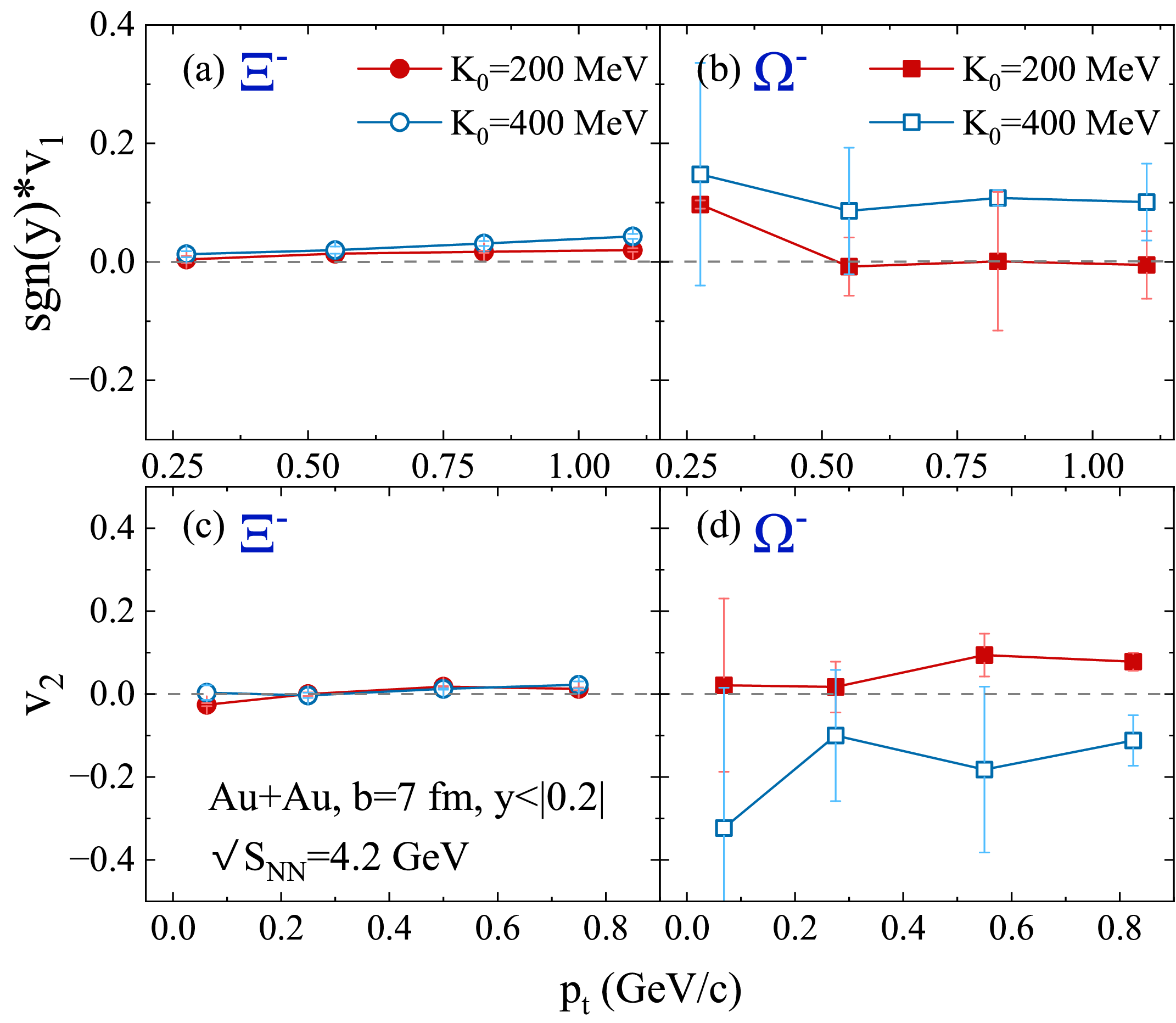}
\caption{Directed flow $sgn(y)*v_{1}$ (upper panels) and elliptic flow $v_2$ (lower panels) as a function of transverse momentum $p_t$ for $\Xi^-$ (left panels) and $\Omega^-$ (right panels) with soft and stiff EoSs in semi-central Au+Au collisions at $\sqrt{S_{NN}}$=4.2 GeV. For both $v_1$ and $v_2$, the rapidity region is limited $y<|0.2|$.} \label{ptv1v2}
\end{figure}
The relationship between the slope of $v_1$ with respect to rapidity ${\mathrm{d} v_1}/{\mathrm{d} y}$ at mid-rapidities and the nuclear matter EoS is direct. A potential indicator of the first-order phase transition from hadron to QGP matter is the minimal value of ${\mathrm{d} v_1}/{\mathrm{d} y}$ as it varies with beam energy. Figure~\ref{v1v2y} (a) showcases $dv_1/dy|_{y=0}$ of $\Omega^-$ and $\Xi^-$ evaluated with both soft and stiff EoSs. Although the general stiff EoS yields a steeper slope of $v_1$ and soft EoS incites a gentler directed flow slope, the treble strangeness $\Omega^-$ exhibits a more potent response to the EoS than the double strangeness $\Xi^-$. This results in an uptick in $dv_1/dy|_{y=0}$ with increasing strangeness content under a stiff EoS and a slide under a soft EoS. Moreover, as illustrated by Figure~\ref{v1v2y} (b), for the mean value $\left \langle v_2 \right \rangle$ of the elliptic flow $v_2$ within $y<|0.2|$, the treble strangeness $\Omega^-$ continues to display greater responsiveness than $\Xi^-$, though lower than that for $dv_1/dy|_{y=0}$. In the present study, $\bar{K}\Xi\rightleftharpoons\Omega\pi$ and, since $\bar{K}$ and $\Xi$ are influenced by the EoS, it's apparent that the sensitivity of $\Omega^{-}$ is in fact a superposition of the sensitivities of $\bar{K}$ and $\Xi$ to the EoS. Therefore, it's predictable to observe that the treble strangeness $\Omega^{-}$ is generally more sensitive to the EoS than the double strangeness $\Xi^{-}$. Whether the slope of $dv_{1}/dy$ for $\Omega^{-}$ is greater or lesser than that of $\Xi^{-}$ is determined by the $\bar{K}$'s sensitivity to the EoS (in the place where $\Xi^{-}$ is produced), which will be evaluated in greater depth later.

The creation of collective flow in non-central heavy-ion collisions is ascribed to the conversion of initial spatial anisotropy into momentum anisotropy \cite{bha2015}. To enhance the visibility of collective movements, semi-central Au+Au collisions at $\sqrt{S_{NN}}=4.2$ GeV were chosen, mirroring the methods of preceding studies. In order to capture more dynamic data and confirm the sensitivity of $\Omega^-$ to the EoS, we've graphed $sgn(y)*v_{1}$ versus the transverse momentum, $p_t$. This is because the directed flow exhibits symmetry with respect to the rapidity distribution, marking distinct traits more effectively. Additionally, elliptic flows $v_2$ are also included for comparison. These simulations are detailed in Figure~\ref{ptv1v2}. It's evident that in comparison to double strangeness $\Xi^{-}$, the transverse momentum distributions of $sgn(y)*v_{1}$ and $v_2$ for the treble strangeness $\Omega^-$ demonstrate their notable sensitivity to the EoS in the transverse momentum ranges of 0.25 to 1 GeV and 0 to 0.8 GeV respectively. Meanwhile, the double strangeness $\Xi^-$ shows less sensitivity to the fluctuations in $sgn(y)*v_{1}$ and $v_2$ induced by the EoS. As a stiffer EoS derives a higher nuclear pressure than a soft EoS, stronger $sgn(y)*v_{1}$ and $v_2$ of the treble strangeness $\Omega^-$ are perceptibly exhibited.

\section{Conclusions}
In conclusion, this study has examined the double strange hyperon $\Xi^-$ and the treble strange hyperon $\Omega^-$ generated in Au+Au collisions at $\sqrt{s_{NN}}$ = 4.2 GeV. It was discovered that the treble strange hyperon $\Omega^-$ demonstrates considerable sensitivity to the EoS than the double strange hyperon $\Xi^-$ in terms of its multiplicity, rapidity distribution, and its directed and elliptic flows. Consequently, the notably peculiar hyperon, $\Omega^-$, can be employed as a valuable gauge for investigating the softening point and potential phase transition of the dense nuclear matter produced in relativistic heavy-ion collisions at pertinent facilities worldwide.

%
The authors thank Zhigang Xiao for initial discussions. This work is supported by the National Natural Science Foundation of China under Grant No. 12275322 and
the Strategic Priority Research Program of Chinese Academy of Sciences with Grant No. XDB34030000.

\end{document}